\documentclass [prl, twocolumn, a4paper]{revtex4}
\usepackage[pdftitle={Dynamics of spin and orbital phase transitions in YVO3},
colorlinks=true,
breaklinks=true,citecolor=blue,pdfkeywords={ultrafast, optically induced phase transition, orbital reordering, phase change, perovskite, orbital order, spin melting, symmetry breaking, nonthermal phase, yttrium vanadate, symmetry relation},a4paper=true]{hyperref}

\usepackage{graphicx}
\usepackage{bm}
\usepackage{dcolumn}
\usepackage{natbib}
\usepackage{color}
\usepackage{amssymb}

\begin{document}

\title{Dynamics of spin and orbital phase transitions in YVO$_3$}

\author{Dmitry A. Mazurenko}
\email{D.A.Mazurenko@rug.nl}
\author{Agung A. Nugroho}
\author{Thomas T. M. Palstra}
\author{Paul H. M. van Loosdrecht}
\email{P.H.M.van.Loosdrecht@rug.nl}
\affiliation{Zernike Institute for Advanced Materials, University of
Groningen, Nijenborgh 4, 9747 AG Groningen, The Netherlands}

\date{\today}

\begin{abstract}
YVO$_3$ exhibits a well separated sequence of orbital and spin order transitions at $200$~K and $116$~K, followed   by a  combined spin-orbital reorientation  at $77$~K. It is shown that   the
spin order can be destroyed  by a sufficiently strong optical pulse within less than $4$~ps. In contrast, the orbital reordering transition from $C$-type to $G$-type orbital order is slower than $100$~ps and goes via an intermediate nonthermal phase.
We propose that the  dynamics of  phase transitions is subjected to symmetry relations between the associated phases.
\end{abstract}


\maketitle
Dynamics of phase transitions is an important and rapidly growing area of modern science. The interest in this topic is not only limited to condensed matter community, but also touches other areas of science ranging from geology \cite{Wookey05} to  modeling of traffic dynamics \cite{Helbing01}.  At the same time, the exponential growth of publications in this area is also driven by applications in optical switching  \cite{myAPL} and phase-change data storage \cite{Lankhorst05} where throughput  is ultimately limited by the phase-change rate. It has been demonstrated that some optically induced structural \cite{Sokolowski95,Rousse01,SokolPRB00,Cavalleri01}, charge \cite{Fiebig01,Ogasawara01}, orbital \cite{Tomimoto03}, and spin order  phase transitions \cite{Beaurepaire96,Hohlfeld97,Stamm07}  may occur on pico- and  sub-picosecond time scales, much faster than phonon-phonon or even electron-phonon equilibration times.  Up to now  research on  the dynamics of so-called nonthermal  phase transitions has mainly been focused on  single phase changes. This paper address  the question on how and which  phase transitions can occur in materials that allow  several phase changes.

An excellent material to pursue this goal is the perovskite YVO$_3$, which shows a sequence of
orbital and spin transitions as a function of temperature.
At room temperature this crystal is paramagnetic and has an  orthorhombic $Pbnm$ structure. At $T_{OO}=200$~K, the
material undergoes a second order transition into an orbitally ordered phase, in  which  the $d_{xz}/d_{yz}$ orbital occupation alternates in all three crystallographic directions  ($G$-type ordering) \cite{Ren98,Blake02}. This transition is accompanied by  a symmetry lowering into a $P2_{1}/b11$ monoclinic form \cite{Tsvetkov04} with a continuous expansion along the $b$ axis and a contraction along the $a$ and $c$ axes \cite{Marquina05}. Below another second order transition at $T_{N} = 116$~K the spins  form an  antiferromagnetic arrangement in the $ab$ plane with a ferromagnetic arrangement along the $c$ axis ($C$-type ordering). Finally,  at $T_{CG} = 77$~K, YVO$_3$  undergoes a first-order phase transition, changing the spin ordering from $C$- to $G$-type with a simultaneous change in the orbital ordering from $G$- to $C$-type.
 This phase transition is accompanied by an abrupt magnetization reversal along the $a$~axis \cite{Ren98,Ren00} and, quite unusually, increases the  symmetry \cite{symcomment} of the lattice back to the orthorhombic $Pbnm$ form \cite{Blake02,Urlich03,Tsvetkov04,Miyasaka03,Miyasaka06}.
All three  phase transitions manifest themselves as  pronounced changes in the  optical absorption in three spectral bands associated with $d-d$ transitions  and located at $1.8$~eV, $2.4$~eV, and $3.3$~eV \cite{Tsvetkov04}. This allows  for tracing the dynamics of the phase transitions by monitoring the optical reflectivity \cite{Tsvetkov04}.
Recently,   optically induced nonthermal melting of the orbital ordering was reported in LaVO$_3$ \cite{Tomimoto03}. This material has a phase diagram similar to YVO$_3$.  An important difference is that  $T_{CG}$ and $T_{OO}$ in LaVO$_3$ are so close to each other that the different phases are not distinguishable in optically induced experiments.
This  Letter reports a study of optically induced phase transitions of the orbital and spin order in
YVO$_3$. In particular, it is shown that  optically induced  spin melting and orbital reordering occur on different time scales, which is ascribed to a difference in symmetry changes
during the phase transitions.

Time resolved two color pump-probe experiments have been performed on $bc$-oriented polished platelets of an YVO$_3$ single crystal  placed in a helium-flow cryostat. The details of the sample growth  can be found in Ref.~\cite{Tsvetkov04}. The optical pump and probe  pulses were derived from an amplified Ti-sapphire laser with a repetition rate of $1$~kHz, in a combination with an optical parametric amplifier. In order to ensure  quasi-homogeneous excitation of the sample, the pump wavelength was set to $800$~nm ($1.55$~eV), which is slightly below the band gap of YVO$_3$. For probing, a wavelength of  $630$~nm ($1.97$~eV) was chosen, since at this wavelength
the reflectivity is sensitive to all the phase transitions.

\begin{figure} [t]
\includegraphics {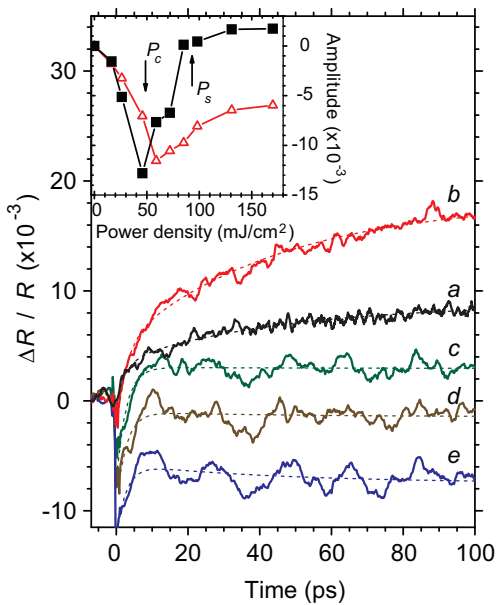}
\caption{\label{fig1:kinet} (Color online).
Dynamics of the transient  reflectivity of YVO$_3$
following  optical excitation with a power density of ($a$) 25~mJ/cm$^2$, ($b$) 45~mJ/cm$^2$, ($c$) 85~mJ/cm$^2$, ($d$) 100~mJ/cm$^2$, and ($e$) 170~mJ/cm$^2$. Solid lines show the experimental data measured at 25~K, dotted lines are a fit of a double exponential function to the experimental data.
The inset shows the amplitude of the  fast  (\textcolor {red}{$\bigtriangleup$}) and slow  ($\blacksquare$) decay components of the transient reflectivity extracted from the fits as a function of the excitation power density. }
\end{figure}

Figure~\ref{fig1:kinet} displays some typical time traces of the
transient reflectivity for different excitation power densities at $T=25$~K, i.e. in the $C$-type orbitally ordered phase.
All  traces show an abrupt change just
after arrival of the pump pulse followed by a kinetics with
pronounced temporal oscillations in the transient reflectivity with
a period of about $19$~ps \cite{TsvetkovNote}.
These oscillations are the result of the interference of the probe
light reflected from the sample surface and from the acoustic strain
wave formed   by an optically induced stress
\cite{Thomsen86}. This
acoustic wave   propagates in a direction perpendicular to the
surface with the velocity $s=\lambda _{probe} / 2 n \tau _{os}
$, with  $n$  the refractive index of YVO$_3$, $\lambda _{probe}$  the
wavelength of the probe light, and $\tau _{os}$ the oscillation
frequency in the reflectivity. Using our experimental data $\tau
_{os}=19$~ps
(Fig.~\ref{fig1:kinet}) and knowing \cite{Tsvetkov04}
$n=2.3$ at $\lambda _{probe}=630$~nm we are able to estimate $s=7.2
\pm 0.5$~km/s, which is close to the longitudinal sound velocity measured in another vanadate perovskite \cite{Maekawa06}. Further  investigation of the acoustical properties of the YVO$_3$
are outside the scope of the present research and, hereafter, our discussion will be focused on
the non-oscillation part of the transient reflectivity dynamics, which
can be quite faithfully fitted with a double-exponential decay
function with $\tau _1=2.7 \pm 1$~ps  and $\tau _2=45 \pm 10$~ps for  fast and slow decay terms, respectively.

For weak optical excitation [Fig.~\ref{fig1:kinet}, curves $a$ and $b$] the slow decay dominates the transient reflectivity dynamics, at late times the transient reflectivity is positive and rises with increasing the excitation power density. However,  when the excitation power density, $P$, overcomes  a critical power density, $P_c$, the slow decay component suddenly disappears  [Fig.~\ref{fig1:kinet} curves $c$-$e$]. Moreover, the saturation level for the  transient reflectivity now becomes more and more negative for stronger excitation. We found that $P_c$ is virtually independent of the initial lattice temperature, as long as $T<T_{N}$, but
depends strongly on the excitation wavelength and varies from
$P_c=55$~mJ/cm$^2$  to $P_c=3$~mJ/cm$^2$ for the pump wavelength set to $800$~nm  and
$400$~nm, respectively. This is most probably due to a substantial change in the
absorption varying by about an order of magnitude in this spectral range \cite{Tsvetkov04}.
The drastic change in the transient reflectivity dynamics at $P=P_c$  is vivid in the inset to Fig.~\ref{fig1:kinet}, which  plots the amplitudes of transient reflectivity decay components as a function of the pump power density. Upon  increasing the excitation power density  the amplitude of the slow decay (solid squares)  grows linearly up to $P=P_{c}$, after which it rapidly decreases until it vanishes at $P_s=1.7 P_c$. In contrast, the amplitude of the fast decay (open triangles) remains quite sizeable even for $P>P_s$.

\begin{figure} [t]
\includegraphics {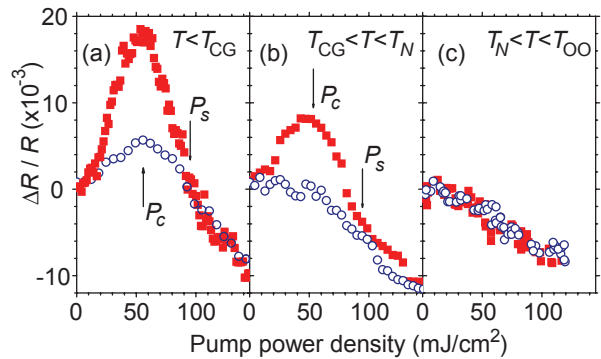}
\caption{\label{fig2:power} (Color online).
Dependence of the transient reflectivity on the pump power density probed
at $4$~ps (\textcolor{blue}{$\circ$}) and
at $100$~ps (\textcolor {red}{$\blacksquare$}) after the optical excitation at  (a) $T=50$~K, (b) $T=100$~K, and (c) $T=140$~K.}
\end{figure}

\begin{figure} [t]
\scalebox {1 }{\includegraphics {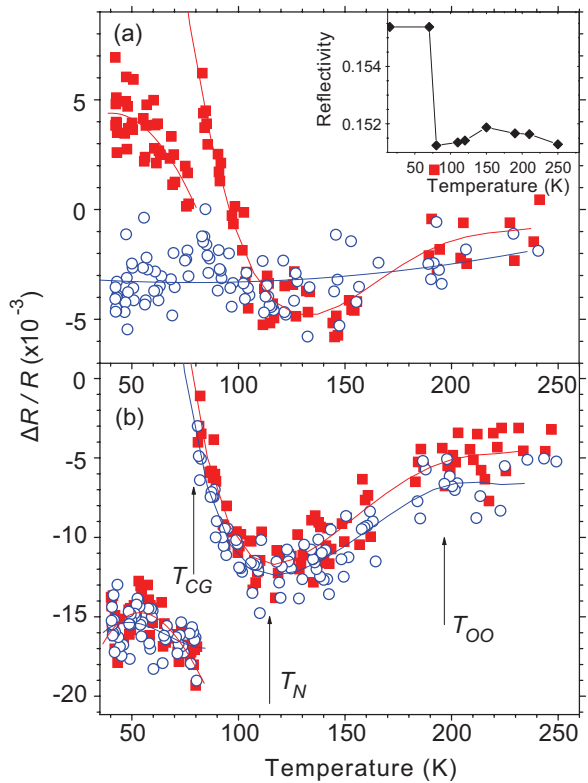}}
\caption{\label{fig3:temper} (Color online).
Temperature dependence of the transient reflectivity  probed
at $4$~ps (\textcolor{blue}{$\circ$}) and $100$~ps (\textcolor{red}{$\blacksquare$}) for (a) $85$~mJ/cm$^2$ and  (b) $190$~mJ/cm$^2$ excitation power density. Lines are guided for eye. Insert shows the stationary reflectance calculated using the optical constants taken from Ref.~\cite{Tsvetkov04}.} \end{figure}

   The observed threshold behavior provides  direct evidence for a \emph{photoinduced phase transition}.  In order to assign this transition to a specific phase change the power dependence of the transient reflectivity has been measured starting out from the three different phases of YVO$_3$: the $C$-type orbital ordered  phase at $T=50$~K [Fig.~\ref{fig2:power}(a)],  the $G$-type orbital ordered phase with $C$-type spin order at $T=100$~K [Fig.~\ref{fig2:power}(b)], and   the $G$-type orbital ordered phase with disordered spins  at $T=140$~K [Fig.~\ref{fig2:power}(c)]. The data are presented for two important delays: at  $4$-ps delay, just after the    fast $\tau _1$-dynamics ends (open circles), and at  $100$-ps delay, when the transient dynamics is over and the reflectivity has  reached  a plateau (solid squares).
   The photoinduced phase transition manifests itself as a   kink at $P=P_c$ [Fig.~\ref{fig2:power}(a,b)], which appears below $T_N$ only.
  Quite importantly, the  shape and the threshold in the power dependence does not change when the sample temperature crosses  $T_{CG}$. However,  when the temperature  approaches  $T_N$ from below, the kink in the power dependence  diminishes   [Fig.~\ref{fig2:power}(b)] and finally disappears at  $T=T_N$. For  $T>T_N$ the transient reflectivity  has a monotonic linear dependence on  the excitation power density [Fig.~\ref{fig2:power}(c)].  The temperature dependence of the transient  reflectivity is  plotted in Fig.~\ref{fig3:temper} for two important  excitation regimes: (a)   $P<P_s$, and (b)  $P>P_s$. In the high excitation regime these dependencies show a kink at $T=T_N$  both for $4$~ps and for $100$~ps delays. In the weak excitation regime the kink in the temperature dependence is notticable only in the $100$-ps trace, but at the same temperature as in the strong excitation regime confirming that the background sample temperature is not affected by the optical excitation.

The disappearance of the  threshold power behavior  above $T=T_N$ indicates that the observed photoinduced phase transition is related to the \emph{melting of the spin order}. This conclusion rises an interesting question on whether the optically induced phase is thermodynamically favorable (thermal) or not. The answer to this question may be deduced from a comparison of the optically  induced changes in  the reflectivity and the variation of the reflectivity observed at different thermal phases. The stationary   reflectivity of YVO$_3$ is plotted  in the inset to  Fig.~\ref{fig3:temper}(a). First, we compare the reflectivity changes for $T>T_{CG}$.  Here, the stationary reflectivity increases in the spin ordered  and decreases in the spin disordered states with elevated temperature. Hence,  the thermally induced change in the reflectivity is positive in the spin ordered ($T_{CG}<T<T_{N}$) and negative in the spin disordered states ($T_{N}<T<T_{OO}$). In turn, for $P<P_c$, the optical excitation of the spin ordered state  induces  a positive change in the reflectivity at 100~ps delay [Fig.~\ref{fig2:power}(b), solid squares]. However, for $P>P_s$, at which the optical excitation destroys the spin order, the transient reflectivity  becomes negative. Further, the transient reflectivity of  the spin disordered state is negative for any excitation power densities [Fig.~\ref{fig2:power}(c)].
  The thermally-driven changes in the reflectivity thus coincide  with the   optically induced changes  measured at $100$~ps delay [Fig.~\ref{fig2:power}(b) and (c), solid squares]   signifying   that at $100$~ps after the optical excitation the spins and electronic subsystems has reached the thermodynamic equilibrium.

 We now turn to the discussion over  the transient reflectivity below the spin-orbital reorientation transition. This transition manifests itself by a substantial decrease in reflectivity [inset to Fig.~\ref{fig3:temper}(a)]. However, the optically induced changes in reflectivity are positive up to the spin melting threshold [Fig.~\ref{fig2:power}(a), $P<P_c$] . This indicates that at $T<T_{CG}$  the arrangement of spins and orbitals are far from  thermal equilibrium even  $100$~ps after the optical excitation.

 The reported experimental data suggest the following model.  In our experiment the pump pulse excites  the electronic  transition from the oxygen $2p$ band to the empty states in the vanadium $3d$ band. Subsequently, these electrons relax to  lower energy states and their excess energy excites the spin and orbital degrees of freedom. For $T>T_N$ the change in the  reflectivity of YVO$_3$ is governed by an arrangement of the orbital occupation, which thermalizes within a time $\tau _1$. In addition, at $T<T_N$,  the photoexcited hot  electrons may also  transfer their energy to  the ordered spin network. At  $P<P_c$, the subsequent thermalization of the spins manifests itself as the slow ($\tau _2$) component in the transient reflectivity traces. The increased  spin temperature and excited orbitals influence reflectivity in opposite directions that leads to  a kink at $T=T_N$ in the temperature (Fig.~\ref{fig3:temper}) and power dependencies [Fig.~\ref{fig2:power}(a) and (b)] of the transient reflectivity.   Another consequence of the rapid disordering of the spin network is disappearance of the slow decay component in the reflectivity kinetics   for  $P>P_s$,  suggesting that  the optically induced melting of the spin order is nonthermal.
 The dynamics of the spin-orbital reordering transition is radically different.
 As it has been mentioned, below the first order phase transition the arrangement of the orbitals remains to be far from the equilibrium even $100$~ps after the optical excitation. However, even in this  initial phase a strong optical excitation may still promote a melting of the spin order in  less than $4$~ps. Indeed, for $P>P_s$ the transient reflectivity measured at $4$~ps and $100$~ps delays coincide with each other [Fig.~\ref{fig3:temper}(b)]. This leads to a conclusion  that the $C\rightarrow G$ orbital transition does not occur on a picosecond time scale and  YVO$_3$ undergoes to a metastable phase with disordered spins but $C$-ordered orbitals.

We suggest that the striking difference in the transition dynamics of the  spin and orbital reordering  and the  spin disordering transitions is related to  their  symmetry change. The rapid spin melting leads to an increase in the spin network symmetry. Oppositely, the spin and orbital reordering is accompanied by a lowering of the lattice symmetry, which requires a considerable time to synchronize the new ordering across the  excited volume. More generally, a  symmetry rule that restricts the speed of phase transition kinetics can be formulated as follow: \emph{The transition rate from a phase with symmetry group $\alpha$  to a phase with symmetry group $\beta$  is limited  if $\alpha$  is  not a subgroup of $\beta$} ( $\alpha \not \subseteq \beta$)
and the limiting factor for  breaking a symmetry is the time required to pass the information about the new arrangements over the excited volume, e.g. it is limited by the propagation velocity of phonons for a structural change.  We suggest that the proposed symmetry rule may be valid for any kind of phase transition unless the excitation creates a coherent wave that breaks  the symmetry itself. To the best of our knowledge,  all reported experimental observations thus far are in agreement with this rule. Indeed, up-to-date  reported ultrafast solid-to-liquid \cite{Sokolowski95,Rousse01,SokolPRB00}, solid-to-solid \cite{Cavalleri01,Fiebig01,myAPL}, charge \cite{Ogasawara01}, orbital \cite{Tomimoto03}, and spin  transitions \cite{Beaurepaire96,Hohlfeld97} are accompanied by $\alpha \subseteq \beta$ symmetry changes. Oppositely, the  phase transitions that obey $\alpha \not\subseteq \beta$ symmetry changes, e.q. amorphous-to-crystalline transition in GeSb films \cite{Callan01}, paraelectric-to-ferroelectric  transition in tetrathiafulvalene-$p$-chloranil \cite{Collet03}, and the spin transition in an organometal spin-crossover material \cite{Huby04}   were reported to be slow.

In conclusions, we demonstrated that the dynamics of the photoinduced  spin disordering transition in $YVO_3$ has a power threshold and occurs on a time scale faster than $4$~ps. In contrast, the orbital reordering transition from $C$-type to $G$-type orbital order was not observed on the picosecond time scale suggesting that it is slower than $100$~ps. We suggest that the difference in the dynamics of these phase transitions may be explained by their symmetry change and proposed a `symmetry' rule, which put a general restriction on  phase transition dynamics.
This rule may prove to have some useful implication, i.e. it limits rates of certain bidirectional phase switching, which is important to consider in designing ultrafast phase change devices.

We acknowledge D.~Fishman for helping with the experiments,  A.F.~Kamp for
technical assistance, and M.H.~Sage for her help in sample preparation. We are grateful to A.A.~Tsvetkov for useful discussion. This work has been supported by the
Netherlands Foundation ``Fundamenteel Onderzoek der Materie''
(FOM),  ``Nederlandse Organisatie voor Wetenschappelijk
Onderzoek'' (NWO), and  by the Royal Dutch Academy of Sciences (KNAW) through the SPIN program.

\bibliographystyle{apsrev}
\bibliography{MazurenkoYVO3_bib}

\begin{thebibliography}{30}
\bibitem[{sym()}]{symcomment}
\bibinfo{note}{The YVO$_3$ lattice may slightly deviate from the assigned
  symmetries as discussed in Refs.~\cite{Tsvetkov04,Blake07}}.
\end{thebibliography}

\end{document}